\def\b{\bibitem}
\documentstyle[aps,prl,epsf,floats]{revtex} 

\def\twocolfiguresize{\ifpreprintsty 17cm \else 17cm \fi}
\begin{document}
% Macros for the various macro package names, etc.
\def\SNG{{\em Physical Review Style and Notation Guide}}
\def\LUG {{\em \LaTeX{} User's Guide \& Reference Manual}}
\def\btt#1{{\tt$\backslash$\string#1}}%
\def\REVTeX{REV\TeX}
\def\AmS{{\protect\the\textfont2
        A\kern-.1667em\lower.5ex\hbox{M}\kern-.125emS}}
\def\AmSLaTeX{\AmS-\LaTeX}
\def\BibTeX{\rm B{\sc ib}\TeX}
%\makeatletter
%\tighten
\twocolumn[\hsize\textwidth\columnwidth\hsize\csname@twocolumnfalse%
\endcsname
\title{First Order Transitions and Multicritical Points in Weak Itinerant 
       Ferromagnets}
\author{D. Belitz\cite{byline1} and T.R. Kirkpatrick\cite{byline2}}
\address{Institute for Theoretical Physics, University of California,
          Santa Barbara, CA 93106}
\author{Thomas Vojta}
\address{Institut f{\"u}r Physik, TU Chemnitz, D-09107 Chemnitz, FRG}
\date{\today}
\maketitle
\begin{abstract}
It is shown that the phase transition in low-$T_c$ clean itinerant ferromagnets
is generically of first order, due to correlation effects that lead to a 
nonanalytic term in the free energy. A tricritical point separates the line 
of first order transitions from Heisenberg critical behavior at higher 
temperatures. Sufficiently strong quenched disorder suppresses the first 
order transition via the appearance of a critical endpoint. A semi-quantitative
discussion is given in terms of recent experiments on MnSi, and predictions 
for other experiments are made.
%
%% 551 characters
%
\end{abstract}
\pacs{PACS numbers: 75.20.En; 75.45.+j; 64.60.Kw } 
]
%\narrowtext
The thermal paramagnet-to-ferromagnet transition at the 
Curie temperature $T_C$ is usually regarded as a prime example
of a second order phase transition.
For materials with high $T_C$ this is well established 
both experimentally and theoretically.
Recently there has been a considerable interest
in the corresponding {\em quantum} phase transition of itinerant
electrons at zero temperature ($T=0$), and in the related finite
$T$ properties
of weak itinerant ferromagnets, i.e. systems with a very low $T_C$.
Experimentally, the transition in the weak ferromagnet
MnSi has been tuned to different $T_C$ by applying hydrostatic
pressure\cite{Lonzarich}. Interestingly, the transition at low $T$
was found to be of {\em first order}, while at higher transition temperatures
it is of second order\cite{heli_footnote}. 
The tricritical temperature that separates the two types
of transitions was found to roughly coincide with the location of a
maximum in the magnetic susceptibility in the paramagnetic phase.
Theoretically, it has been shown \cite{us_dirty,us_clean} that in a $T=0$
itinerant electron system, soft modes that are unrelated to the critical
order parameter (OP) or magnetization fluctuations couple to the latter.
This leads to an effective long-range 
interaction between the OP fluctuations. In disordered systems, the additional 
soft modes are the same `diffusons' that cause the so-called weak-localization 
effects in paramagnetic metals\cite{LeeRama}. In clean systems
there are analogous, albeit weaker, effects that manifest
themselves as corrections to Fermi liquid theory\cite{us_fermions}.
A Gaussian theory is sufficient to obtain 
the exact quantum critical behavior in the most interesting dimension, $d=3$,
for clean as well as for disordered systems (apart from logarithmic corrections
in the clean case)\cite{us_dirty,us_clean}.

In this Letter we show that at sufficiently low 
temperatures the phase transition in
itinerant ferromagnets is {\em generically} of first order. This surprising 
result is shown to be rooted in fundamental 
and universal many-body physics underlying the 
transition, viz. long-wavelength correlation effects,
and hence to be independent of the band structure.
This suggests that the behavior observed in MnSi is
generic, and should also be present in other weak itinerant ferromagnets.
We also make detailed predictions about how quenched disorder suppresses 
the first order transition, which allows for
decisive experimental checks of our theory.

Let us start by deriving the functional form of the free energy
of a bulk itinerant ferromagnet at
finite $T$, and in the presence of quenched disorder that we
parametrize by $G=1/\epsilon_{\rm F}\tau$, with $\epsilon_{\rm F}$ 
the Fermi energy, and $\tau$ the elastic mean-free time. The general
Landau expansion of the free energy $F$ as a function of the magnetic
moment $m$ in an approximation that neglects OP fluctuations is
\begin{mathletters}
\label{eqs:1}
\begin{equation}
F = t\,m^2 + u_4\,m^4 + u_6\,m^6 + \ldots\quad.
\label{eq:1a}
\end{equation}
The coefficients $t$, $u_4$, $u_6$, etc. in this expansion can have 
nontrivial properties and contain important physics. 
A derivation from a microscopic theory shows that they
are given as frequency-momentum integrals over
correlation functions in a `reference system' that depends on the nature
of the underlying microscopic model\cite{Hertz}. If the critical magnetization
fluctuations are the only soft modes in the system, then they
are simply numbers. However, if in the process of
deriving the Landau functional some other soft modes have been integrated
out, then the coefficients will in general not exist, since they are
represented as diverging integrals over the soft modes. In
Refs.\ \onlinecite{us_dirty} and \onlinecite{us_clean} it was shown that 
in an itinerant electron system at $T=0$ there are indeed such soft modes.
In the disordered case, these
are the `diffusons' mentioned above, with a dispersion relation 
$\omega\sim k^2$, and they lead to coefficients whose
divergent parts have the form
\begin{equation}
u_{2m} \propto \int_{0}^{\Lambda} dk\,k^2 \int d\omega\ 
                \frac{1}{(\omega + k^2)^{2m}}\quad.
\label{eq:1b}
\end{equation}
\end{mathletters}%
Here $\Lambda$ is a momentum cutoff, and all prefactors in the integrals
have been omitted. In the clean case, the relevant
soft modes are particle-hole excitations in the spin-triplet channel with
a ballistic dispersion relation, $\omega\sim k$. The resulting integrals
are still divergent, although not as strongly as in the disordered case.
It was shown in Refs.\ \onlinecite{us_dirty},\onlinecite{us_clean} that
these divergent terms in the Landau expansion can be understood as an
illegal expansion of a nonanalytic term in the free energy of the form
\begin{equation}
f(m) = m^4 \int_{0}^{\Lambda} dk\,k^2 \int_{0}^{\infty} d\omega\ \frac{(-1)^x}
       {\left[(\omega + k^x)^2 + m^2\right]^2}\quad.
\label{eq:2}
\end{equation}
In the disordered case, where $x=2$, this follows explicitly from
Eq.\ (3.6') of Ref.\ \onlinecite{us_dirty}. In the clean case, an analogous 
treatment yields the same expression with $x=1$.
Notice the different sign of the dirty case 
compared to the clean one, which we will 
come back to below. Equation (\ref{eq:2}) yields
$f(m) \propto m^{5/2}$ and $f(m) \propto m^4\ln m$ in
the disordered and clean cases, respectively.
In either case the resulting singularity is protected by the magnetization,
which gives the soft modes a mass. The leading effect of $T\neq 0$
is adequately represented by replacing $\omega \rightarrow
\omega + T$. In addition, in the presence of disorder
the ballistic modes in the clean case obtain a mass proportional to $1/\tau$,
so the appropriate generalization of 
Eq.\ (\ref{eq:2}) for the clean case ($x=1$) 
to finite temperature and disorder
is obtained by the replacement $\omega \rightarrow \omega + T + 1/\tau$. 
Doing the integrals, and adding the usual terms of order
$m^2$ and $m^4$, we obtain a free energy of the form
\begin{eqnarray}
F&=&t\,m^2 + G\,(N_{\rm F}\Gamma_t)\,m^4\,
     \left[m^2 + (\alpha T)^2 \right]^{-3/4}
\nonumber\\
 &&+ v\,m^4\,\ln\left(m^2 + (T + \beta G)^2\right) + u\,m^4 + O(m^6)\ ,
\label{eq:3}
\end{eqnarray}
where $\Gamma_t$ is an effective spin-triplet interaction 
amplitude\cite{us_dirty} made
dimensionless by means of a density of states at the Fermi level, $N_{\rm F}$.
If we measure $F$, $m$, and $T$ in terms of a microscopic energy, e.g.
$\epsilon_{\rm F}$, then $t$, $v$, and $u$ are all dimensionless.
$v$ is quadratic in $\Gamma_t$\cite{us_clean}.
$t=1-N_{\rm F}U$ is the dimensionless
distance from the critical point. It depends on the physical spin-triplet
interaction amplitude $U$, with $N_{\rm F}U \approx 1$
in a ferromagnetic or nearly ferromagnetic system, while $\Gamma_t$ 
above is an effective interaction amplitude with 
$N_{\rm F}\Gamma_t < 1$. 
$\Gamma_t$ is expected to be relatively larger in strongly correlated
systems. Finally, $\alpha$ and $\beta$ are
parameters that measure the relative strengths of the temperature and the
disorder dependence, respectively, in the two nonanalytic terms.
They are numbers of order unity, and like $u$ and $v$ they are non-universal.
Equation\ (\ref{eq:3}) provides a functional form of the free energy that
correctly describes the leading nonanalytic $m$-dependence for both clean
and disordered systems, as well as the leading temperature cutoff for 
either term and the leading disorder cutoff for the clean nonanalyticity.

The sign of $v$ merits some attention.
Perturbation theory to second order in $\Gamma_t$ yields 
$v>0$\cite{us_clean,us_chi_s}. Further,
$v>0$ indicates a decrease of the effective Stoner coupling
constant $I$ due to correlation
effects: $I$ is a homogeneous spin
susceptibility, $v>0$ means that this susceptibility
increases as the wavenumber increases from zero\cite{us_chi_s}, and
correlation effects decrease with increasing wavenumber. It is well known that
correlation effects in general decrease $I$\cite{White}, and $v>0$ is
consistent with that. Ref.\ \onlinecite{us_clean}
has given some possible mechanisms for $v$ to be
negative at least in some materials, and showed that in this case the 
ferromagnetic transition is always of second order.
However, the generic case is $v>0$, which we will now discuss.

\begin{figure}[t]
\epsfxsize=60mm
\centerline{\epsffile{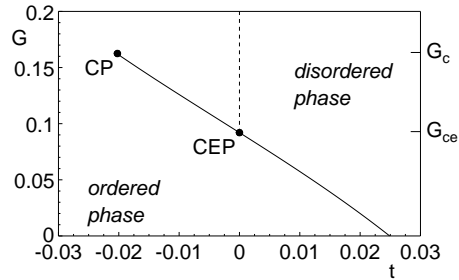}}
\caption{Phase diagram at $T=0$ for $u=1$, $v=0.5$, 
 $N_{\rm F}\Gamma_t=0.5$,
 $\alpha=\beta=1$, showing a second order transition
 (dashed line), and a first order transition (solid line).}
\label{fig:1}
\end{figure}
We first consider
the case $T=0$. The transition in the clean system, $G=0$, is then of first 
order, since $m^4\ln m <0$ for small $m$. Upon disordering the system, 
$G>0$, the negative term is no longer the leading one at $t=0$. For small
values of $G$, the transition remains first order. However, for
$G$ exceeding a value $G_{ce}$ the first order transition occurs only at
$t<0$, and it is pre-empted by a second order transition. Since the
negative term is only the third term in an $m$-expansion of $F$, the
multicritical point where the nature of the transition changes is
a critical endpoint (CEP)\cite{ChaikinLubensky}. 
The phase diagram in the $G$-$t$ plane
is shown in Fig.\ \ref{fig:1}. For
$G_{ce} < G < G_c$, the second order transition at $t=0$ is followed by a
second transition, the second one being of first order, to a state with
a larger magnetization. The line of first order transitions ends
in a critical point (CP) at
a disorder value $G_c$, where the two minima in the free energy merge. 

Before we consider $T>0$, let us discuss this result and the validity of
our conclusions. To facilitate an analytic discussion, we put $\beta=0$.
We then have
$F = tm^2 + G(N_{\rm F}\Gamma_t)m^{5/2} + 2vm^4\ln m + um^4$. 
At $G=0$ there is
a first order transition at $t=v\exp [-(1+u/v)]$, and the magnetization
at the transition has a value $m=\exp [-(1+u/v)/2]$. Notice that the 
nonanalytic term is the {\em leading} one in $F$ after the $tm^2$ term, and
that we know the functional form of $F$ {\em exactly} up to $O(m^4)$.
As long as $u/v>>1$, $m$ is exponentially small at the transition. For
small $v$, our Landau expansion is therefore controlled in the sense
that terms of $O(m^6)$ and higher would have to have exponentially large
coefficients in order to change our results. 
For $G>G_{ce}=(4v/3N_{\rm F}\Gamma_t)\exp[-(1+3u/4v)]$, the first 
order transition is pre-empted
by a second order one. At the CEP, the magnetic
moment has the value
$m=\exp [-(2/3 + u/2v)] = e^{-1/6}\,m(G=0)$. Allowing for $\beta = O(1)\neq 0$,
and repeating the calculation numerically,
leads only to minor quantitative changes of these results.
\begin{figure*}[ht]
\epsfxsize=\twocolfiguresize
\centerline{\epsffile{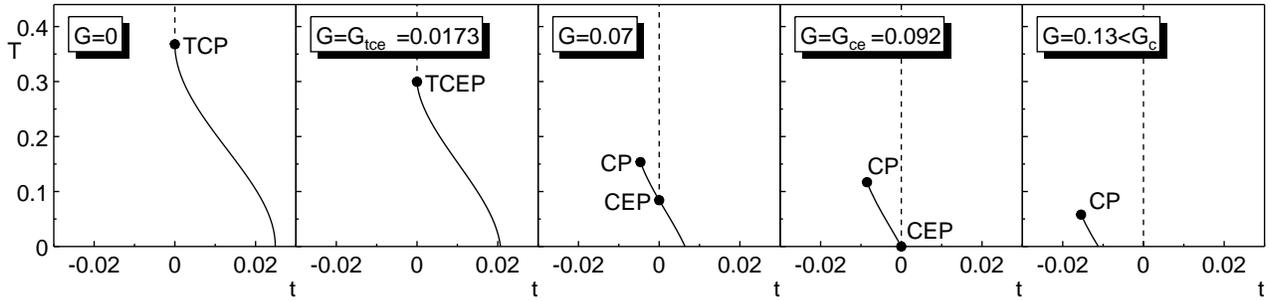}}
\caption{Phase diagrams for $u=\beta=1$, $v=\alpha=N_{\rm F}\Gamma_t=0.5$
 showing first (solid) and second (dashed) order transitions.}
\vskip -3mm
\label{fig:2}
\end{figure*}

At $T>0$, the free energy is an analytic function of $m$, but for
small $T$ the coefficients in an $m$-expansion become very large. Our
remarks about the validity of our truncated Landau expansion therefore
still apply, i.e., at $0<k_{\rm B}T<<\epsilon_{\rm F}$, 
our theory contains the most
important terms to every order in an expansion in powers of $m^2$.
Let us first consider the clean system, $G=0$. There is a tricritical point
(TCP) at $T_{\rm tc} = \exp (-u/2v)$, with a first order
transition for $T<T_{\rm tc}$, and a line of Heisenberg critical points for
$T>T_{\rm tc}$. To describe the (conventional) tricritical behavior
in $d=3$ our mean-field theory is sufficient (apart from logarithmic 
corrections)\cite{tricriticality}, for the critical behavior at $T>T_{\rm tc}$
it is of course not.

\begin{figure*}[ht]
\epsfxsize=\twocolfiguresize
\centerline{\epsffile{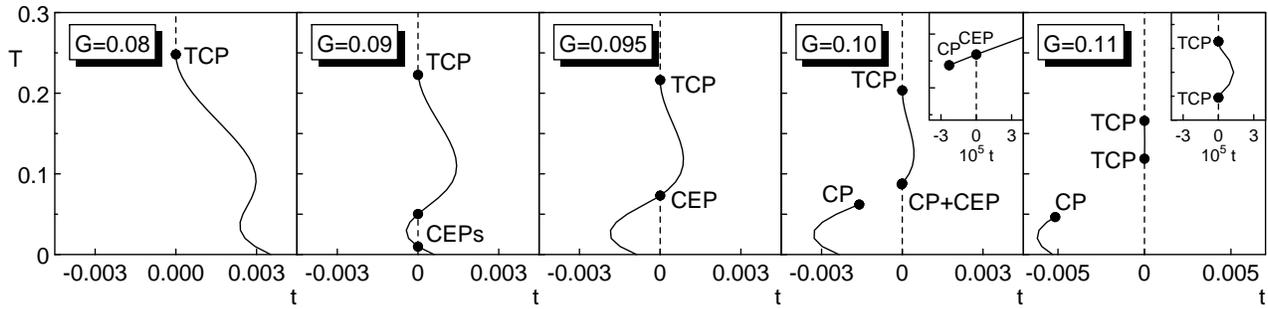}}
\caption{Same as Fig.\ \ref{fig:2}, but for $\alpha = 2$.}
\vskip -3mm
\label{fig:3}
\end{figure*}

For the suppression of the first order transition by disorder at $T>0$
we find two different possibilities,
depending on the value of the parameter $\alpha$.
For small $\alpha$ ($\alpha\alt 1.5$ with our choice of the
remaining parameters, Fig.\ \ref{fig:2}), the TCP is replaced by a CEP
for $G$ larger than some $G_{tce}<G_{ce}$. At $G=G_{ce}$, the CEP
reaches $T=0$, and for larger values of $G$ the transition is of second
order for all $T$. At small $T$, it is followed by a first order
transition. The line of first order transitions ends in a critical point,
and disappears only for $G=G_c$. For larger values of 
$\alpha$ (Fig.\ \ref{fig:3}), the TCP persists for a range of disorder
larger than $G_{ce}$. The first order transition first gets pre-empted
in a temperature window between two CEPs. At $G=G_{ce}$, the lower
CEP reaches $T=0$, while the TCP at higher temperature survives.
With further increasing disorder, two CPs appear in the
ordered phase, and the remaining CEP gets replaced by a TCP. Finally,
the two TCPs merge, and the remaining CP reaches $T=0$, 
eliminating the last temperature regions with first order
transitions.
Notice that the interesting features of these phase
diagrams do not depend on the logarithm in Eq.\ (\ref{eq:3});
similar features are obtained in standard phenomenological Landau
expansions with a negative coefficient of the third term\cite{BruceCowley}.
We stress again, however, that in our case the expansion is controlled, and
that we have a definite physical mechanism for the 
appearance of a negative term,
in contrast to purely phenomenological theories.

We now turn to a discussion of the available experimental information
on this subject. MnSi has a low $T_C$
($\approx 30\,{\rm K}$) under ambient pressure, and $T_c$ can be driven to
zero by a hydrostatic pressure $p_c\approx 15 {\rm kbar}$. 
$k_{\rm B}T/\epsilon_{\rm F}<<1$ always, and $T$ is low enough to suppress
phase breaking processes, so the quantum critical
behavior is easily accessible experimentally. This system
has been studied in detail by Pfleiderer et al.\cite{Lonzarich} These
authors found from susceptibility measurements that the transition turns 
first order at a $T_c$ of about 12 K. The line of second order transitions
was found to scale with pressure like $T_c \propto (p_c-p)^{3/4}$, while
in the first order regime the transition temperature varies like
$T_1\propto (p-p_c)^{1/2}$.
The scaling of $T_c$ with pressure was explained by a
scaling analysis based on the self-consistently renormalized (SCR) theory of
Moriya and Kawabata\cite{Moriya}, assuming a dynamical 
exponent $z=3$. The first order
transition at low $T$ was attributed in Ref.\ \onlinecite{Lonzarich} 
to a sharp structure in the density of states at the Fermi 
level.

Let us look at the experiment in the light of the above 
discussion. In 
Ref.\ \onlinecite{us_clean} it was shown that the quantum phase transition in
$d=3$ is indeed correctly described by SCR theory, 
apart from logarithmic corrections that would
be very difficult to detect experimentally, and that the dynamical critical
exponent in $d=3$ is $z=3$. The analysis of 
Ref.\ \onlinecite{Lonzarich} was therefore adequate, and in particular the
quantum-to-classical crossover exponent $\phi$, which determines the behavior
of the critical temperature as a function of $t$ through the relation
$T_c\propto t^{\phi}$, has a value $\phi=3/4$. If one makes the plausible
assumption that $t$ depends linearly on the hydrostatic pressure, at least
for small $t$, then this is in agreement with both the experimental 
finding and the analysis in Ref.\ \onlinecite{Lonzarich}. As
for the pressure dependence of $T_1$,
one of the temperature scales in the problem is the
Fermi liquid temperature scale\cite{us_clean}, 
which arises from a quadratic $T$-dependence 
of $t$. Since the first order transition is determined by the 
condition $t(T_1)={\rm const.}$, we immediately get $T_1 \propto \sqrt{p_c-p}$,
where we again assume a linear relation between $p$ and $t$. 

We finally discuss the observation\cite{Lonzarich}
that the tricritical temperature roughly coincides with a minimum of the
inverse magnetic susceptibility 
$\chi^{-1}$ in the paramagnetic phase.
In $d$ dimensions, the leading $T$-dependence of the 
paramagnetic susceptibility is of the form\cite{us_chi_s}
\begin{equation}
\chi/2N_{\rm F} = 1 + 2{\tilde v}_d\,T^2\,T^{d-3} - {\tilde u}_d\,T^2
                    \quad,
\label{eq:4}
\end{equation}
In $d=3$, the nonanalyticity is of the form $T^2\ln T$. A calculation of
${\tilde v}_3$ to second order in $\Gamma_t$
revealed\cite{us_chi_s} that to that order, ${\tilde v}_3 = 0$,
in agreement with prior results from Fermi liquid 
theory\cite{CarneiroPethick}. Ref.\ \onlinecite{us_chi_s}
also discussed that there are reasons to believe that the {\em exact}
value of ${\tilde v}_3 = 0$ may be nonzero. If we assume that this is 
the case, then we obtain a minimum in $\chi^{-1}$ at
a temperature $T_{\rm min} = \exp(-{\tilde u}_3/2{\tilde v}_3 - 1/2)$.
Since the nonanalyticities in $F$ and $\chi$ are manifestations of the
same singularity, one expects ${\tilde u}_3 \approx u$ and
${\tilde v}_3 \approx v$, so that $T_{\rm min} \approx T_{\rm tc}$.
While this provides a possible explanation for the observation, we stress 
the speculative nature
of the above considerations due to the theoretical uncertainty about 
a nonanalytic $T$ dependence of $\chi$ in $d=3$.

Our theory thus provides us with a complete explanation for the nature of
the transitions observed in MnSi, and in particular for the existence of
a first order transition at low $T$, which in Ref.\ \onlinecite{Lonzarich}
was attributed to a band structure feature characteristic of MnSi. While
this feature may well be sufficient to make the transition in MnSi of first
order, the present theory leads to the surprising prediction that the first
order transition is {\em generic}, and thus should be present in other weak
itinerant ferromagnets as well. Our theory further predicts in detail how the
first order transition will be suppressed by quenched disorder. Observations
of such a suppression, or lack thereof, would be very interesting for
corroborating or refuting the theory. Semi-quantitatively, the theory
predicts that the $T$ region that shows a first order transition will
be largest for strongly correlated systems.
Conversely, since the dependence of the
tricritical temperature on the system parameters is exponential,
in some, or even many, systems the first order transition may take place
only at very low temperatures. This may explain why in ZrZn$_2$
no first order transition has been observed\cite{Lonzarich}, although the
experiment does not seem to rule out a weakly first order
transition\cite{Pfleiderer}.

We gratefully acknowledge helpful conversations with G. Lonzarich and 
C. Pfleiderer. This work was
supported in part by the NSF under grant Nos. PHY94--07194,
DMR--98--70597, and DMR--96--32978, and by the DFG under grant No. SFB 393/C2.

\vfill\eject
\end{document}